\documentclass[reprint,
superscriptaddress,11ptamsmath
 amsmath,amssymb,
 aps,
prl,
floatfix,
]{revtex4-2}

\usepackage{physics}
\usepackage{cases}

\usepackage[small]{caption}
\usepackage{dcolumn}
\usepackage{bm}

\ifx\pdfoutput\undefined
\usepackage[dvipdfmx]{graphicx}
\usepackage[dvipdfmx]{hyperref}
\else
\usepackage{graphicx}
\usepackage{hyperref}
\fi

\usepackage{txfonts}
\usepackage{braket}

\begin{document}

\title{
    Many-body Keldysh Crossover in the DC-driven Haldane Spin Chain
}

\author{Koichi Okazaki}
\affiliation{The Institute for Solid State Physics, The University of Tokyo, Kashiwa, Chiba 277-8581, Japan}
\author{Shun Okumura}
\affiliation{Department of Applied Physics, The University of Tokyo, Hongo, Bunkyo-ku, Tokyo 113-0033, Japan}
\author{Shintaro Takayoshi}
\affiliation{Department of Physics, Konan University, Kobe, 658-8501, Japan}
\author{Takashi Oka}
\affiliation{The Institute for Solid State Physics, The University of Tokyo, Kashiwa, Chiba 277-8581, Japan}

\date{\today}

\begin{abstract}
    We theoretically study nonlinear processes driven by a DC spin-electric field in the antiferromagnetic spin-1 Heisenberg model starting from the ground state in the Haldane phase.
    The DC spin-electric field generates finite spin current and accumulation since the symmetry protected topological order is destroyed by the field.
    We find two microscopic mechanisms responsible for the breakdown and a crossover between them:
    In weak fields, tripron-antitripron pair creation occurs through the tunneling mechanism, and in strong fields, the system is described by an effective Hamiltonian breaking the protecting symmetries.
    We analyze the numerically obtained results in terms of the Dykhne--Davis--Pechukas theory and Floquet theory, verifying the universal picture of the many-body Keldysh crossover in the DC-driven quantum spin systems.
\end{abstract}

\maketitle

\textit{Introduction}:
Many-body effects and topology are fundamental concepts in condensed matter physics.
The study of ground state properties in many-body systems has significantly deepened our understanding of how these systems can be characterized.
Beyond the traditional classification criteria based on symmetry breaking, such as in ferromagnets and superconductors, topological criteria have also emerged, revealing the existence of topologically ordered states.
These states can exhibit both short- and long-range entanglement, as exemplified by the Haldane phase in the $S=1$ Heisenberg model~\cite{AKLT} and the fractional quantum Hall state, respectively.
Research of the excited states of topologically ordered systems is also underway~\cite{CASPERS1982103, AROVAS1989431, Moudgalya2018}.

In addition to equilibrium properties, nonequilibrium properties in many-body systems have garnered significant interest due to advances in experimental techniques such as ultrafast pump-probe spectroscopies~\cite{Iwai2006, Giannetti2016, Basov2017, TorreRMP}.
Intense coherent fields can drive many-body systems out of their ground states.
For instance, in strongly correlated electronic systems, applying THz-electric fields can induce a transition from an insulating state to a metallic one~\cite{Mayer2015, Yamakawa2017, Li2022}.
This process typically begins with pair creation of charge carriers~\cite{Oka2003} induced by quantum tunneling~\cite{Landau, Zener1932, Stuckelberg}.
As the frequency of the AC field increases or the amplitude decreases, the excitation mechanism shifts from tunneling to multi-photon absorption, which is known as the Keldysh crossover~\cite{Keldysh1965, Oka2012}, and has been experimentally demonstrated in Mott insulators recently~\cite{Li2022}.

Nonequilibrium properties of many-body topological systems are less understood, with a few exceptions such as the breakdown of topological order~\cite{TakayoshiPRB2014}, generation of chiral spin order~\cite{Claassen2017, Kitamura2017}, and the Floquet topological ordered phase~\cite{PoPRX2016}.
In these latter examples, systematic exploration was conducted using Floquet theory for quantum states in time-periodic Hamiltonians~\cite{EckardtRevModPhys.89.011004, Oka2019, Harper2020}.
Although, tunneling breakdown and Keldysh crossover have been studied in many-body systems~\cite{Oka2003, Oka2012,Mayer2015, Yamakawa2017, Li2022}, research on how they can affect a topologically ordered ground state is still not understood, especially in interacting spin systems.

\begin{figure*}[tb]
    \centering
    \includegraphics[width=\hsize]{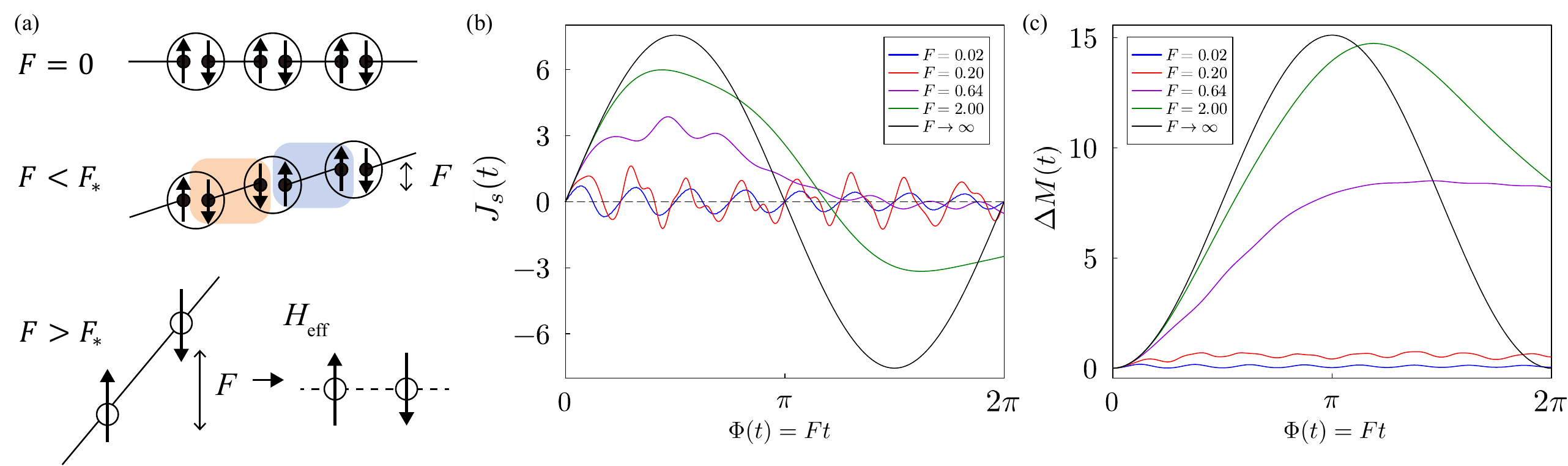}
    \captionsetup{justification=raggedright}
    \caption{
        (a) Schematic pictures of states of the $S=1$ Heisenberg model driven by a DC spin-electric field $F$ in Eqs.~\eqref{eq:Hamgrad} and \eqref{eq:SpinHam}.
        For $F=0$ and $F<F_*$, states are represented with $S=1/2$ spins depicted as short arrows in the circle using the valence bond solid picture.
        Triplon-antitriplon pairs depicted as orange and blue shaded bonds are created for $0<F<F_*$.
        In strong fields $F>F_*$, the state is described by the effective Hamiltonian $H_{\mathrm{eff}}$ in Eq.~\eqref{eq:Heff} for $S=1$ spins (long arrow).
        (b), (c) Time evolution of the (b) spin current $J_{\mathrm{s}}(t)$ and (c) the magnetization $\Delta M(t)$ for various $F$.
    }
    \label{fig:Fig1}
\end{figure*}

In this work, we consider the dynamics of the Haldane chain in the presence of a DC spin-electric field.
In lattice models, the Hamiltonian becomes time-periodic in DC-fields as the periodicity is given by the Bloch oscillation.
Thus, we expect a DC-version of the Keldysh crossover to take place as shown in Fig.~\ref{fig:Fig1}~(a).
In the weak field limit, we find the creation of {\it triplon-antitriplon pairs}, where the transition probability is compared with the Dykhne--Davis--Pechukas (DDP) theory~\cite{Dykhne, Davis}.
On the other hand, in the strong field limit, we obtain the effective Floquet Hamiltonian using the Floquet--Magnus (FM) expansion of the time evolution operator with $1/F$ as the small expansion parameter.
In both mechanisms, the nonequilibrium ecxitaions are viewed as a topological breakdown accompanying with breaking of time-reversal and space-inversion symmetries by the DC spin-electric field.

\textit{Spin-1 Heisenberg model in a spin-electric field}:
We consider the spin-1 antiferromagnetic Heisenberg model with a gradient magnetic field:
\begin{align}
    \hat{H} & = \frac{J}{2}\sum_{j} 
    \left[\left(\hat{S}_j^+\hat{S}_{j+1}^- +\mathrm{H.c.}\right)
        +2\hat{S}_{j}^z\hat{S}_{j+1}^z\right]
    -F\sum_{j}
    j\hat{S}_j^z,
    \label{eq:Hamgrad}
\end{align}
where $\hat{S}_j^\mu$ ($\mu=x,y,z$) is the spin-1 operators at the site $j$, and $\hat{S}_j^\pm = \hat S_j^x\pm i\hat S_j^y$.
The coupling constant $J>0$ is the antiferromagnetic exchange interaction, and $F$ is the gradient of the external magnetic field.
Hereafter we take the unit $\hbar=1$ and lattice constant $a=1$. 
We define the twist operator as
\begin{align}
    \hat{g}(\theta)=\exp(-i\theta\sum_{j}j\hat{S}_j^z).
    \label{eq:TwistOperator}
\end{align}
By performing the time-dependent unitary transformation $\hat{g}(Ft)$, the Hamiltonian in Eq.~\eqref{eq:Hamgrad} is recast into
\begin{align}
    \hat H' & = \hat{g}(Ft)\hat{H}\hat{g}^\dagger(Ft)
    -i\hat{g}(Ft)\partial_t \hat{g}^\dagger(Ft)
    \nonumber                                                                                                                            \\
            & = \frac{J}{2}\sum_{j}\left[\left(e^{iFt}\hat S_j^+\hat S_{j+1}^- +\mathrm{H.c.}\right)+2\hat S_{j}^z\hat S_{j+1}^z\right].
    \label{eq:HamTransf}
\end{align}
Note that $Ft$ plays a role similar to the vector potential representing static electric fields in lattice electronic systems, and thus we denote $F$ as a {\it spin-electric field}.

In the following, we consider the model
\begin{align}
    \hat H\left(\Phi(t)\right) = \frac{J}{2}\sum_{j=1}^{L}\left[\left(e^{i\Phi(t)}\hat S_j^+\hat S_{j+1}^- +\mathrm{H.c.}\right)+2\hat S_{j}^z\hat S_{j+1}^z\right]
    \label{eq:SpinHam}
\end{align}
with periodic boundary condition $L+1 \equiv 1$ where $L$ is the number of sites, and $\Phi(t)$ is the $\mathrm{U}(1)$ gauge field for spins.
The ground state of the spin-1 Heisenberg antiferromagnet at $\Phi=0$ is in the valence bond solid phase [Fig.~\ref{fig:Fig1}(a)]~\cite{Haldane1983,AKLT}, which is a gapped symmetry protected topological phase~\cite{Pollmann2012}.
We study the time evolution of this system starting from the ground state with $\Phi=0$ as an initial state.
The DC spin-electric field is switched on at $t=0$, i.e., $\Phi(t)=Ft$ ($t>0$). The time evolution is governed by the time-dependent Schr\"odinger equation
\begin{align}
    i\partial_{t}\ket{\psi(t)}=\hat{H}(\Phi(t))\ket{\psi(t)}.
    \label{eq:Time_dep_Sch}
\end{align}
We numerically obtain the initial state by exact diagonalization and integrate Eq.~\eqref{eq:Time_dep_Sch} using the fourth-order Runge-Kutta method.
In the calculations below, we set the system size $L=8$ and take the antiferromagnetic exchange interaction $J=1$ as the unit of energy.

\textit{Spin current and induced magnetization}:
As noted above, $\Phi$ behaves as a gauge field for the spins.
Therefore, we can define the spin current operator as
\begin{align}
    \hat J_{\mathrm{s}}(\Phi(t))
    = \frac{d}{d\Phi}\hat{H}(\Phi(t))
    = \sum_{j=1}^L \hat{j}_{\mathrm{s},j}(\Phi(t)),
\end{align}
where $\hat{j}_{\mathrm{s},j}(\Phi(t))=\frac{iJ}{2}\left(e^{i\Phi(t)}\hat S_j^+\hat S_{j+1}^- - e^{-i\Phi(t)}\hat S_j^-\hat S_{j+1}^+\right)$ represents the local spin current density.
Conservation law $d\braket{\hat{S}_j^z}/dt=i\braket{[\hat{H}(\Phi(t)),\hat{S}_j^z]}
    =\braket{\hat{j}_{\mathrm{s},j-1}(\Phi(t))-\hat{j}_{\mathrm{s},j}(\Phi(t))}$ holds in this system due to the rotational symmetry of the Hamiltonian around the $z$-spin axis, and thus the spin current is a well-defined physical quantity.

We plot the time evolution of the spin current $J_{\mathrm{s}}(t) = \braket{\psi(t) | \hat{J}_{\mathrm{s}}(\Phi(t)) | \psi(t)}$ in Fig.~\ref{fig:Fig1}(b) and the magnetization (spin accumulation) $\Delta M(t) = \int_{0}^{t} J_{\mathrm{s}}(t')dt'$ in Fig.~\ref{fig:Fig1}(c) for various strengths of the spin-electric field $F$.
We numerically find that they behave qualitatively different in the weak and strong field limits as follows.

\begin{figure}[tb]
    \centering
    \includegraphics[width=0.8\hsize]{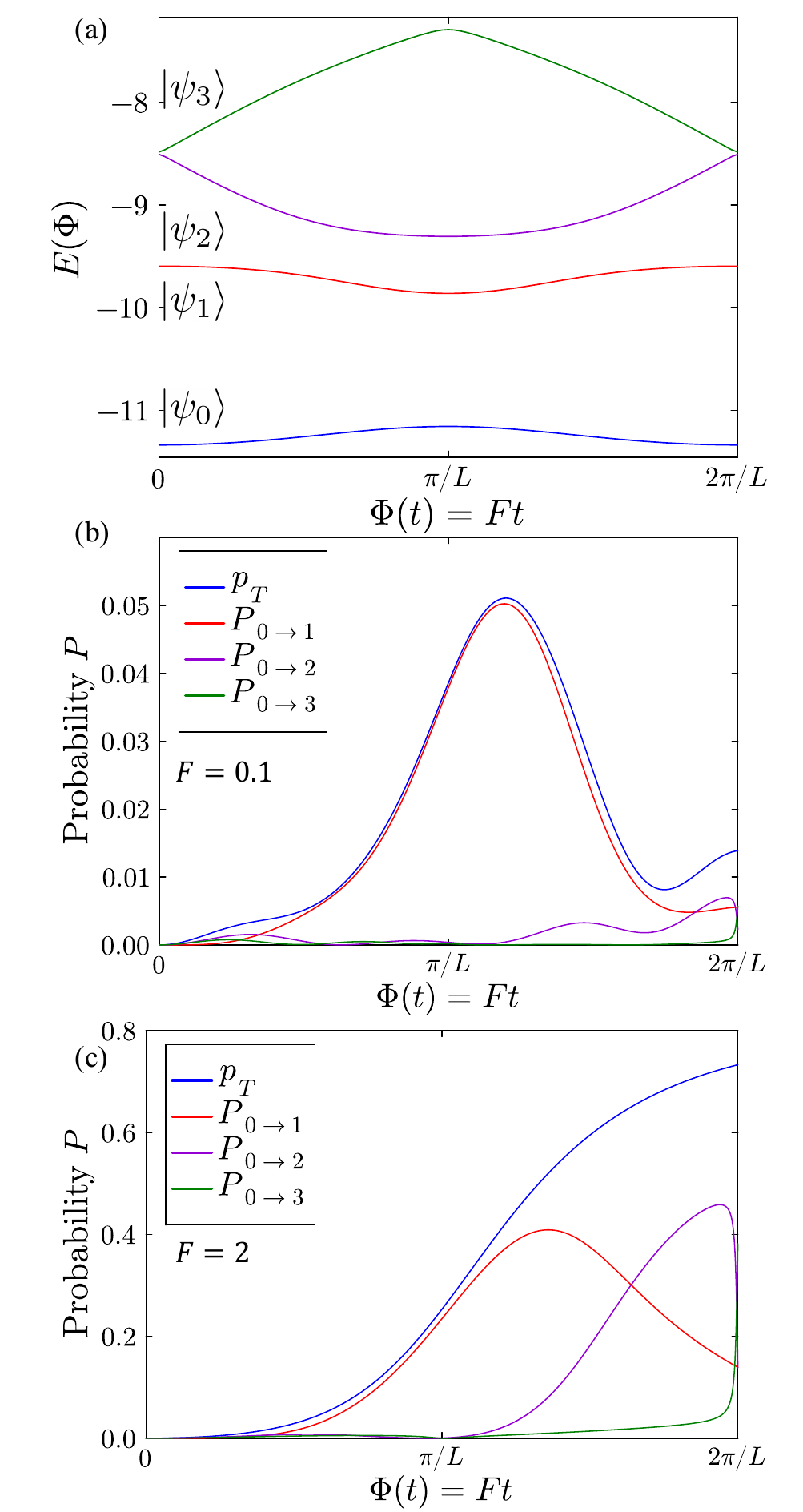}
    \captionsetup{justification=raggedright}
    \caption{
        (a) Instantaneous eigenenergy spectrum plotted against the gauge field $\Phi(t)$.
        (b)(c) Time dependence of the transition probabilities to the $n$-states $ P_{0\to n}$ and the total transition probability $p_\mathrm{T}$ for spin-electric fields (b)$F=0.1$ and (c)$F=2$.
    }
    \label{fig:Fig2}
\end{figure}

    {\bf Weak field}:
Both $J_\mathrm{s}(t)$ and $\Delta M(t)$ oscillate around zero with a period approximate to $Ft=2\pi/L$.
This is the near adiabatic regime where the dynamics reflect the periodicity of the instantaneous energy spectrum of $\hat{H}(\Phi(t))$ (see Fig.~\ref{fig:Fig2} and following section).

    {\bf Strong field}:
In the large $F$ limit, the state does not evolve from the initial state since the time duration becomes small.
Still, there is finite spin current since the spin current operator is time-dependent leading to $J_{\mathrm{s}}(t)\to A\sin(\Phi(t))$, where $A = 2J\left|\bra{\psi_0(0)}\sum_{j}\hat S_j^x\hat S_{j+1}^x\ket{\psi_0(0)}\right|$.
The oscillation has a similar nature as the Bloch oscillation in fermionic systems and the period is given by $T=\frac{2\pi}{F}$.

As $F$ increases from weak to strong limits, the oscillation with the period $2\pi/(FL)$ is smeared out,
and net magnetization increases. For example, if we see the data for $F=2$,
the oscillation with the period $2\pi/(FL)$ vanishes, and another oscillation with the period $2\pi/F$ emerges.
This qualitative change of behaviors implies that there is a crossover as $F$ is increased.
The intriguing behavior of the spin current is qualitatively similar to that of the electric current in the Mott insulator induced by DC-electric fields~\cite{Oka2003}.
It has been argued that the dynamics of electrons are approximately described by quantum tunneling in the weak electric field regime, while DC-Stark localization happens in the strong field limit.

\textit{Energy spectrum and transition probability}:
To investigate the microscopic excitation mechanisms, let us first introduce the instantaneous eigenstates $\ket{\psi_n(\Phi)}$ and eigenenergies $E_n(\Phi)$ satisfying
\begin{align}
    \hat{H}(\Phi)\ket{\psi_n(\Phi)}
    =E_n(\Phi)\ket{\psi_n(\Phi)}.
    \label{eq:instantaneous}
\end{align}
We focus on excited states with zero total $z$-spin and zero total momentum.
The conservation of total momentum is due to the translational symmetry described by $[\hat{H}(\Phi),\hat{T}]=0$, where $\hat{T}$ is the translation operator: $\hat{T}\hat{S}_{j}^\alpha \hat{T}^{-1}=\hat{S}_{j+1}^\alpha$.
We employ the periodic boundary condition leading to $\hat{T}^L=1$ and thus, the total wave number is conserved as $\hat{T}\ket{\psi(t)} = e^{i2\pi l/L}\ket{\psi(t)}$ ($l=0,\;1,\;\ldots,\;L-1$) with $l$ being the total momentum.
If we start from the ground state, $l=0$ is conserved during the evolution.
We depict the instantaneous eigenenergy spectrum in Fig.~\ref{fig:Fig2}~(a).
The ground state $E_0$ is separated from the excited states by a finite energy gap.
Note that this energy gap relevant to our dynamics is not the well-known Haldane gap;
the Haldane gap involves states with finite momentum.

The spectrum has a $\Phi=2\pi/L$-periodicity that leads to the $Ft=2\pi/L$ oscillation of the spin current in the weak field limit.
This periodicity originates from the relation
\begin{align}
    \hat{g}^\dag (2\pi/L)\hat{H}(\Phi+2\pi/L)\hat{g}(2\pi/L) = \hat{H}(\Phi)
\end{align}
induced by the twist operator Eq.~\eqref{eq:TwistOperator}.
The instantaneous eigenstates satisfy $\ket{\psi_n(\Phi+2\pi/L)} = \hat{g}(2\pi/L)\ket{\psi_n(\Phi)}$ which gives $E_n(\Phi+2\pi/L)= E_n(\Phi)$.
In the adiabatic limit, the spin current for the $n$-th eigenstate is given by $J_s(t)=\frac{d}{d\Phi}E_n(\Phi(t))$, which explains its periodic behavior.

Next, we investigate the transition probability defined as
\begin{align}
    P_{0\to n}(t) \equiv
    \left| \braket{\psi_n(\Phi(t))|\psi(t)} \right|^2.
    \label{eq:trans_prob}
\end{align}
We also define the survival probability $P_{0\to 0}(t)$, and the total transition probability as $p_{\mathrm{T}}(t) \equiv \sum_{n\neq 0} P_{0\to n}(t) = 1 - P_{0\to 0}(t)$.
We show the time evolution of $p_{\mathrm{T}}(t)$ and $P_{0\to n}(t)$ ($n=1,2,3$) in the cases of $F=0.1$ and $F=2$ in Figs.~\ref{fig:Fig2}(b) and \ref{fig:Fig2}(c), respectively.
For $F=0.1$, the transition from $\ket{\psi_0(\Phi(t))}$ to $\ket{\psi_1(\Phi(t))}$ mainly occurs around $Ft=\pi/L$, which is precisely where the gap between the two states $\ket{\psi_0(\Phi)}$ and $\ket{\psi_1(\Phi)}$ takes a minimum as seen in Fig.~\ref{fig:Fig2}(a).
This signals that the transition mechanism in the small field region is a many-body version of the Landau-Zener-St\"uckelberg (LZS) transition, which we will quantify below.
In the many-body version, we have many levels and successive tunneling takes place unlike the one-body two-level system~\cite{SM}, as we can see in the next transition occurring from $\ket{\psi_1(\Phi(t))}$ to $\ket{\psi_2(\Phi(t))}$ around $Ft=\frac{3}{2}\pi/L$.
In contrast, for strong fields ($F=2$), the transition probability becomes large and the adiabatic picture breaks down; The state quickly evolves away from the ground state.
We will see below that the large field limit is described by the Floquet theory.

\begin{figure}[tb]
    \centering
    \includegraphics[width=\hsize]{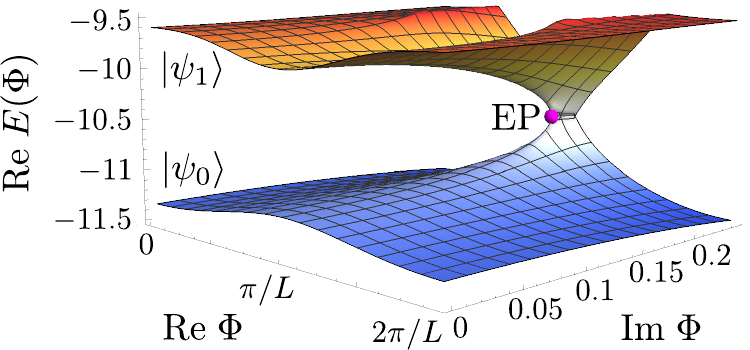}
    \captionsetup{justification=raggedright}
    \caption{
        3D-plot of the real part of eigenenergy spectrum $\mbox{Re}\,E$ for $\hat{H}(\Phi)$ with complex gauge field $\Phi$. The magenta ball represents the position of the exceptional point (EP) where eigenenergies of $|\psi_0\rangle$ and $|\psi_1\rangle$ merge on $\mbox{Re}\;\Phi = \pi/L$.
    }
    \label{fig:Fig3_ep}
\end{figure}

\textit{Weak and Strong field limits and Many-body Keldysh crossover}:
The behaviors of spin current and transition probability indicate that there is a qualitative difference between the weak and strong field regimes.
Now let us examine the mechanisms of excitation in more detail.

\begin{figure}[t]
    \centering
    \includegraphics[width=\hsize]{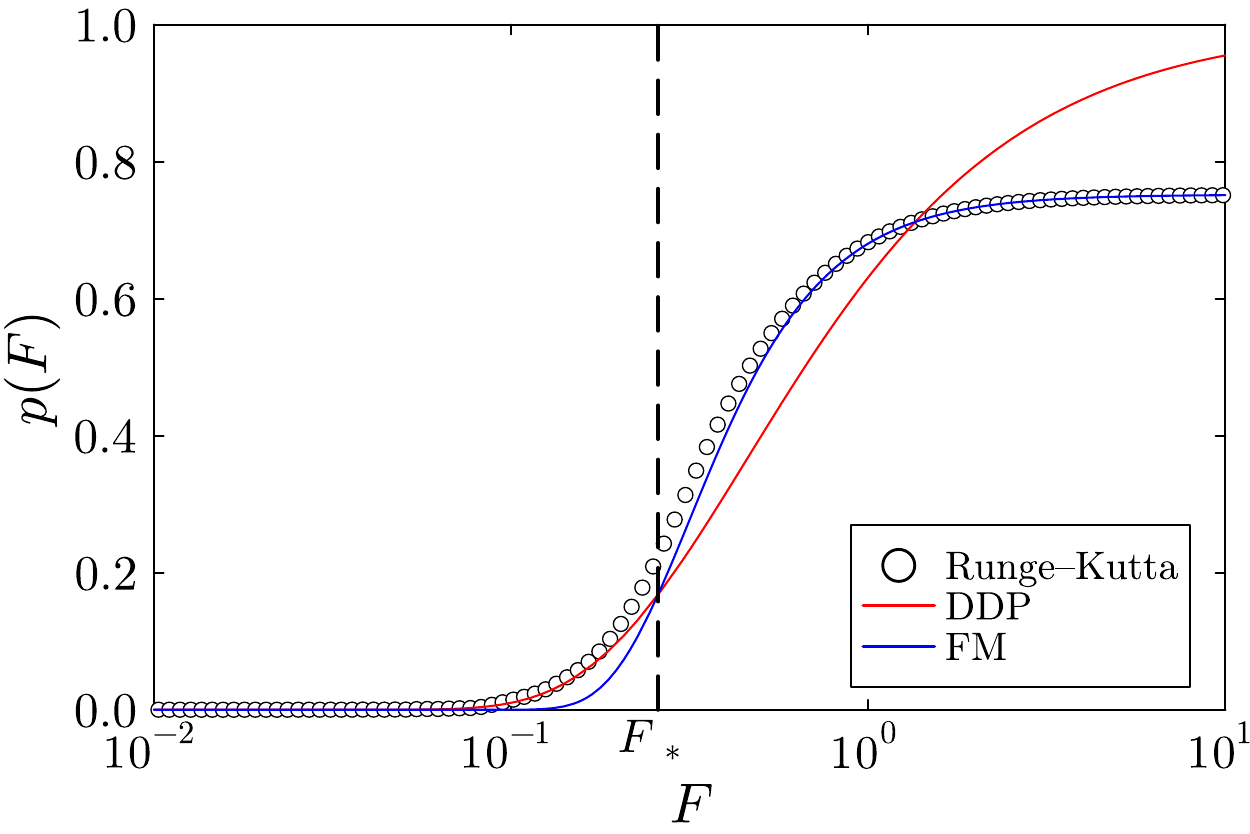}
    \captionsetup{justification=raggedright}
    \caption{
        Spin-electric field $F$ dependence of the tunneling probability. The red line corresponds to the Dykhne--Davis--Pechukas (DDP) method, whereas the blue curve corresponds to the Floquet--Magnus (FM) expansion.
    }
    \label{fig:Fig4}
\end{figure}

    {\bf Weak field}:
For weak fields, we are near the adiabatic limit and the excitation is approximately described as tunneling between the ground and first excited states, i.e., $\ket{\psi_0(\Phi(t))}$ and $\ket{\psi_1(\Phi(t))}$ [Fig.~\ref{fig:Fig2}(b)].
One can study the many-body LZS tunneling by extending the DDP theory to many-body energy levels~\cite{Oka2010, Oka2012}.
In the DDP theory~\cite{Dykhne, Davis}, the transition probability is obtained from the spectral flow of a non-Hermitian Hamiltonian that is obtained by analytically continuing the time variable to imaginary time.
Here, we apply the many-body DDP theory to the Heisenberg model.
To this end, let us extend the parameter $\Phi=Ft$ in Eq.~\eqref{eq:instantaneous} to complex values. 
In Fig.~\ref{fig:Fig3_ep}, we show the extension of Fig.~\ref{fig:Fig2}(a), which plots $\textrm{Re}\;E(\Phi)$ as a fucntion of complex values of $\Phi$.
Along the line for $\Phi=\pi/L+iy$ ($y \in \mathbb{R}$), there exists an exceptional point (EP), where the Hamiltonian becomes nondiagonalizable.
Note that $\Phi=\pi/L$ ($y=0$) corresponds to the minimum of the energy gap.
It is seen that, as $y$ increases, the two eigenvalues $E_0$ and $E_1$ merge at the EP $\Phi_{\mathrm{c}}=\pi/L+iy_{\mathrm{c}}$ and then split into two conjugate eigenvalues for $y>y_{\mathrm{c}}$.
The tunneling probability $p(F) \equiv p_{\mathrm{T}}(2\pi/(LF))$ is calculated by the DDP method to be $p(F)=\exp(-\pi F_\mathrm{th}/F)$~\cite{Oka2010,Oka2012}, where
\begin{align}
    F_\mathrm{th} = \frac{2}{\pi} \mathrm{Im}\int_{\pi/L}^{\Phi_c} d\Phi [E_1(\Phi)-E_0(\Phi)]
    \label{eq:ddp}
\end{align}
and the integration path is $\Phi=\pi/L+iy$ ($0\leq y\leq y_{\mathrm{c}}$).
Equation~\eqref{eq:ddp} can be evaluated by exact diagonalization and numerical integration, and we obtain $F_\mathrm{th}\simeq 0.1465$ for $L=8$.
In Fig.~\ref{fig:Fig4}, we show the tunneling probability $p(F)$ obtained from numerical time evolution by the Runge--Kutta method as circles and the prediction from the DDP method, $\exp(-\pi F_\mathrm{th}/F)$, with $F_\mathrm{th} \simeq 0.1465$ as a red solid line.
They agree well for small $F$, but the deviation increases in the strong field regime.

{\bf Strong field}:
In the strong field region, the behavior of the system can be understood using the FM expansion~\cite{Casas2001,Blanes2009, Eckardt2015, Eckardt2017, Mikami2016} in the Floquet theory.
This is because the Hamiltonian $\hat{H}(Ft)$ in Eq.~\eqref{eq:SpinHam} has a periodicity $\hat H(Ft)=\hat H(Ft+2\pi)$
and the field strength $F$ plays the role of ``frequency".
The time evolution operator $\hat{U}(t)$ involving $\ket{\psi(t)}=\hat{U}(t)\ket{\psi(0)}$ is written as
\begin{align}
    \hat{U}(t) = \mathcal{T}\exp\left(-i\int_{0}^{t}\hat{H}(Ft') dt'\right),
    \label{time_evo}
\end{align}
where $\mathcal{T}$ is the time ordered product.
This operator can be decomposed as $\hat U(t) = e^{-i\hat{\Lambda}(t)} e^{-i\hat{H}_{\mathrm{eff}}t}$, where $\hat{\Lambda}(t)$ is the kick operator.
The Floquet effective Hamiltonian
\begin{align}
    \hat{H}_\mathrm{eff} = J\sum_{j}\left(1+\frac{J}{2F}\left(\hat S_j^z-\hat S_{j+1}^z\right)\right)\hat{S}^z_j\hat{S}^z_{j+1}+\cdots
    \label{eq:Heff}
\end{align}
is obtained by the FM expansion~\cite{Casas2001,Blanes2009, Eckardt2015, Eckardt2017, Mikami2016} with respect to $1/F$.
The lowest order of the $F$-induced terms in \eqref{eq:Heff} contains a three-spin non-reciprocal interaction which breaks both space-inversion and time-reversal symmetries.
Similarly, by expanding the kick operator $\Lambda(t)$ with respect to $1/F$, $\hat{U}(t)$ is expanded as
\begin{align}
    \hat{U}(t) = 1-\frac{i}{F}\hat{u}_1(t)
    -\frac{1}{F^2}\hat{u}_2(t)+O(F^{-3}).
\end{align}
The explicit forms of $\hat{u}_1(t)$ and $\hat{u}_2(t)$ are given in Supplementary Material~\cite{SM}.
Thus, noting that $\ket{\psi_0(2\pi/L)}=\hat{g}(2\pi/L)\ket{\psi_0(0)}$, we derive the survival probability $P_{0\to 0} = |\braket{\psi_0(2\pi/L)|\hat{U}(2\pi/L)|\psi_0(0)}|^2$ as
\begin{align}
    P_{0\to 0} \approx \left|z-\frac{i\langle \hat g^\dag\hat u_1\rangle}{F}-\frac{\langle \hat g^\dag\hat u_2\rangle}{F^2}\right|^2,
    \label{eq_quench}
\end{align}
where $\braket{\cdot}=\braket{\psi_0(0)|\cdot|\psi_0(0)}$, $\hat{g}\equiv\hat{g}(2\pi/L)$, and $z=\braket{\hat{g}^{\dagger}}$.
We obtain the tunneling probability as $p(F)=1-P_{0\to 0} \approx (1-|z|^2)\exp(-F_\mathrm{q}^2/F^2)$, where $F_q^2 = (1-|z|^2)^{-1} (|\braket{\hat g^\dag \hat u_1}|^2-2z\braket{\hat g^\dag\hat u_2})$.
The ground state expectation value of the twist operator $z$ is a $\mathrm{U}(1)$ quantity related to polarization and localization of the many-body wave function\cite{Resta1998, Resta}.
By numerical calculations, we evaluate $\braket{\hat g^\dag \hat u_1} \simeq 4.567$, $\braket{\hat g^\dag \hat u_2} \simeq -20.86$, $z \simeq -0.4982$, and $F_q^2\simeq 0.0998$ for $L=8$.
This prediction of $p(F)$ by the FM expansion is shown as a blue solid line in Fig.~\ref{fig:Fig4}, which agrees well with the result of numerical time evolution by the Runge--Kutta method in the large-$F$ region.
In the $F\to \infty$ limit, $p(F)$ asymptotically approaches $1-|z|^2=1-|\braket{\psi_0(2\pi/L)|\psi_0(0)}|^2$, which corresponds to the case that the Hamiltonian is quenched and the quantum state remains in the initial state.

From the above analyses, we find a crossover between the weak and strong field limits where the transition probability changes as
\begin{numcases}{p(F)=}
    \exp\left(-\pi \frac{F_\mathrm{th}}{F}\right) &
    ($F \lesssim F_*$), \label{eqsLZformula}\\
    (1-|z|^2)\exp\left(-\frac{F_q^2}{F^2}\right) &
    ($F \gtrsim F_*$). \label{Eq_quench}
\end{numcases}
The crossover field $F_*$ can be defined as the smallest solution of $\exp(-\pi F_\mathrm{th}/F_*)=(1-|z|^2)\exp(-F_\mathrm{q}^2/F_*^2)$ where the dominant mechanism switches: $F=F_*$ is the first intersecting point of the red and blue solid lines in Fig.~\ref{fig:Fig4}.
This is the many-body DC-Keldysh crossover in quantum spin systems induced by spin-electric fields which is schematically depicted in Fig.~\ref{fig:Fig1}(a).
In the weak field side, the dynamic is described by tunneling excitation of triplon-antitriplon pairs from the valence bond solid ground state, whereas the strong field side is described by a Floquet-Magnus effective Hamiltonian that breaks time-reversal and space-inversion symmetries.
Both two microscopic mechanisms identified above explain how the symmetry protected topological order of the $S=1$ Heisenberg model is dynamically destroyed by the spin-electric fields from the different viewpoints.

\textit{Conclusion and discussion}:
In summary, we have theoretically investigated the nonlinear excitation caused by a spin-electric field in the Haldane chain.
We have numerically demonstrated that spin current and accumulation of magnetization are induced and their behaviors show a contrast between the weak and strong field regimes.
We have also evaluated the transition probabilities from the ground state to the excited states.
In the weak field regime, the excitation process can be explained in terms of the tunneling process and the tunneling probability is approximately obtained by the DDP method.
The strong field regime can be described by the Floquet theory with inverse frequency expansion.
As the spin-electric field increases, the crossover between these two regions takes place, which means that we have demonstrated the many-body Keldysh crossover in the DC-driven quantum spin chain.

As a future perspective, it would be interesting to perform further investigations on the nonequilibrium dynamics of quantities that are directly connected to topological order such as the string order parameter and entanglement entropy.
Especially, it is worth investigating the relationship between the transition propability and topological gap opening.
Another intriguing future problem is to study topologically ordered systems with long range entanglement such as fractional quantum Hall states.

\section*{ACKNOWLEDGMENTS}
K.O. is supported by the World-leading Innovative Graduate Study Program for Materials Research, Industry, and Technology (MERIT-WINGS) of the University of Tokyo.
This work was supported by a Grant-in-Aid for Scientific Research from JSPS, KAKENHI Grant Nos.~JP23K22418, JP24H00191, JP24K06891, JP23H04865, JP23K22487, JP22K13998, JP23K25816 and JST CREST Grant No.~JPMJCR19T3, Japan.

\bibliography{cite}

\clearpage

\section{Two-level Landau--Zener model}
For simplicity, let us consider the two-level system, i.e., the linearly driven Landau--Zener (LZ) model:
\begin{align}
    \hat H(t)=\begin{pmatrix}
                  vFt & m    \\
                  m   & -vFt
              \end{pmatrix},
\end{align}
where $F$ is a field strength, $2m$ is the energy gap and $v$ is the energy slope.
The conventional LZ problem is how the tunneling probability, from the ground state at $t=-\infty$ to the excited state at $t = +\infty$, is given.
We treat a different time-evolution problem: that starts at $t=-\theta/F$ and ends at $t=+\theta/F$, to obtain the hint to look at many-body problems in which the time evolution occurs within the finite time. We see that the transition probability is no longer the same as the LZ formula in a nutshell. The numerical result for $m=v=1,~\theta = \pi$ is shown in Fig.~\ref{LZprob}(a) (black dots).

\subsection{Small F regime}
It is still well-matched with the LZ formula \cite{Landau, Zener}:
\begin{align}
    p(F) = \exp(-\pi\frac{m^2}{vF})
\end{align}
because $\theta/F$ may be regarded as large as $\infty$ when a field strength is weak.
The LZ formula can be also derived by the Dykhne--Davis--Pechukas (DDP) method \cite{Dykhne, Davis}. Let the eigenvalues of $H(t)$ be $E_\pm(t)$, then the tunneling probability is given by
\begin{align}
    p(F) = \exp\left(-2\mathrm{Im}\int_0^{t_c} dt~(E_+(t)-E_-(t))\right),
\end{align}
where $t_c = im/vF$ is a gap-closing point, i.e., $E_+(t_c)=E_-(t_c)$. The tunneling probability calculated by this formula is plotted in Fig.~\ref{LZprob}(a) (red line).

\subsection{Large $F$ regime}
\begin{figure}[t]
    \centering
    \includegraphics[width=6cm]{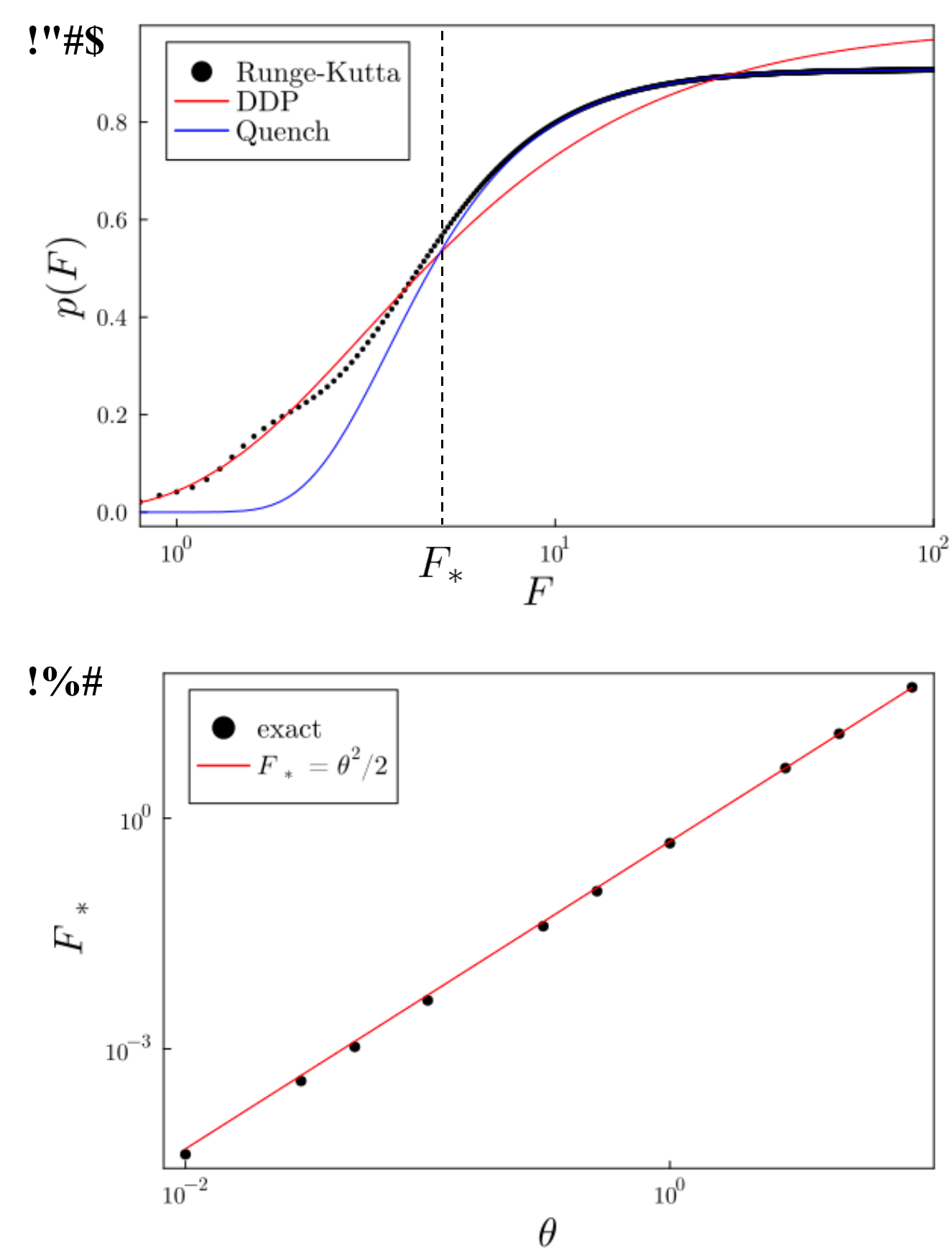}
    \captionsetup{justification=raggedright}
    \caption{(a) $F$-dependence of the transition probability of the LZ model, for $m=v=1,\theta = \pi$. The dots (black) show the numerical result calculated by the 4th-order Runge-Kutta method. The red curve and blue curve show the LZ formula (or DDP formula) and the fitting curve of the quenching limit, respectively. (b) $\theta$ dependence of the turning point $F_*$.}
    \label{LZprob}
\end{figure}

The tunneling probability is approximately given for large $F$ by
\begin{align}
    p(F) \propto \exp(-\frac{4m^2\theta^2}{3F^2}),
\end{align}
which we call the \textit{quenching limit}. The plot of Eq.~(4) as a function of $F$ is shown in Fig.~\ref{LZprob}(a) (blue line).
This type of problem has already been discussed \cite{Vitanov1, Vitanov2}, but we obtain a simple and novel formula expressed by an elementary function.

To prove this new formula, one may evaluate the unitary time-evolution operator,
\begin{align}
    \hat U=\mathcal{T}\exp(-i\int_{-\theta/F}^{\theta/F}dt~\hat H(t)).
\end{align}
One of the possible ways is to expand it to the second order of $1/F$.
It may be expanded as
\begin{align}
    \hat U = & 1-i\frac{2m\theta}{F}\hat \sigma_x-\frac{2m^2\theta^2}{F^2}+i\frac{4mv\theta^3}{3F^2}\hat \sigma_y \nonumber \\
             & -i\frac{2mv^2\theta^4}{F^3}\hat \sigma_z+\frac{(\cdots)}{F^3}\hat \sigma_x+O(1/F^4),
\end{align}
where $\hat \sigma_\alpha$s are the Pauli matrices.
The initial ground state and the final excited state are respectively given by
\begin{align}
    \ket{\psi_{i,0}} & =\frac{1}{\sqrt{2E(E-\theta)}}\begin{pmatrix}
                                                         m \\ -E+\theta
                                                     \end{pmatrix}, \\
    \ket{\psi_{f,1}} & =\frac{1}{\sqrt{2E(E+\theta)}}\begin{pmatrix}
                                                         E+\theta \\ m
                                                     \end{pmatrix},
\end{align}
where $E=\sqrt{m^2+\theta^2}$.
Since the final state is $\hat U\ket{\psi_{i,0}}$, the transition amplitude is easily obtained by Eqs. (6),(7) and (8):
\begin{align}
    \bra{\psi_{f,1}}\hat U\ket{\psi_{i,0}}= & \frac{v\theta}{\sqrt{m^2+v^2\theta^2}}\left(1-\frac{2}{3}\left(\frac{(m\theta)^2}{F^2}\right)\right)\nonumber \\
                                            & -i\frac{2mv^2\theta^4}{9F^3}+O(1/F^4).
\end{align}
Finally, for large enough $F$, the tunneling probability is approximated by
\begin{align}
    p(F) & = \frac{v^2\theta^2}{m^2+v^2\theta^2}\left(1-\frac{(2m\theta)^2}{3F^2}\right)+O(1/F^4)\nonumber \\
         & \simeq \frac{v^2\theta^2}{m^2+v^2\theta^2}\exp\left(-\frac{(2m\theta)^2}{3F^2}\right).
\end{align}
Here we use the approximation, $\exp(-x)\simeq 1-x$ for small $x$.

\subsection{Turning point}

The transition probability $p(F)$ is given by
\begin{align}
    p(F) & =|\braket{\psi_1(t=+\theta/F)}{\Psi(\theta/F)}|^2 \nonumber \\
         & \simeq \begin{cases}
                      \displaystyle \exp(-\pi \frac{m^2}{vF})~(F<F_*), \\
                      \\
                      \displaystyle \frac{v^2\theta^2}{m^2+v^2\theta^2}\exp(-\frac{4m^2\theta^2}{3F^2})~(F>F_*),
                  \end{cases}\label{eqsLZformula_}
\end{align}
where $F_*$ is the turning point defined as the intersection of the quenching limit curve and the LZ formula curve as shown in Fig.~\ref{LZprob}(a). The exact value of the turning point $F_*$ is given by
\begin{align}
    \frac{2v\theta^2}{F_*}=\frac{3\pi}{4}+\sqrt{\frac{3v^2\theta^2}{m^2}\log\left(\frac{
            v^2\theta^2}{m^2+v^2\theta^2}\right)+\frac{9\pi^2}{16}},
\end{align}
but Fig.~\ref{LZprob}(b) tells us that $F_*(\theta)$ is expressed approximately as $F_* \propto \theta^2$.

\section{Floquet--Magnus expansion}

We perform the Floquet--Magnus (FM) expansion with respect to $1/F$ for the Hamiltonian
\begin{align}
    \hat{H}(t) = \frac{J}{2}\sum_{j=1}^{L}\left[\left(e^{iFt}\hat S_j^+\hat S_{j+1}^- +\mathrm{H.c.}\right)+2\hat S_{j}^z\hat S_{j+1}^z\right]
\end{align}
The time evolution operator is written as
\begin{align}
    \hat{U}(t) = \mathcal{T}\exp\left(-i\int_{0}^{t}\hat{H}(Ft') dt'\right),
\end{align}
where $\mathcal{T}$ is the time ordered product.
We decompose this operator as
\begin{align}
    \hat{U}(t) = e^{-i\hat{\Lambda}(t)} e^{-i\hat{H}_{\mathrm{eff}}t}.
\end{align}
Using the Fourier component of the Hamiltonian
$\hat{H}(t)=\sum_{m}\hat{H}_m \exp(-imFt)$, we can express $\hat{\Lambda}^\mathrm{FM}(t)$ and $\hat{H}_{\mathrm{eff}}^{\mathrm{FM}}$ as
\begin{align}
    i\hat{\Lambda}^\mathrm{FM}(t)   & = i\frac{\hat{J}_{\mathrm{s}}(0)-\hat{J}_{\mathrm{s}}(Ft)}{F} \notag                    \\
                                    & \quad +\frac{[\hat{H}(Ft)-\hat{H}(0),\hat{H}_0]+i[\hat{H}_1,\hat{H}_{-1}]\sin(Ft)}{F^2} \\
    \hat H^\mathrm{FM}_\mathrm{eff} & = \hat H_0 - \frac{[\hat{H}_1,\hat{H}_{-1}]+i[\hat{J}_{\mathrm{s}}(0), \hat{H}_0]}{F},
\end{align}
where $\hat{H}_{\pm1}=\frac{J}{2}\sum_j \hat S_j^\mp\hat S_{j+1}^\pm$ and $\hat{H}_0 = J\sum_j \hat S_j^z\hat S_{j+1}^z$. This is

\begin{align}
    \hat U(t)   & = 1-\frac{i}{F}\hat u_1(t) -\frac{1}{F^2}\hat u_2(t)+O(F^{-3}),                                                         \\
    \hat u_1(t) & = \int_0^{Ft}d\Phi\; \hat H(\Phi), \notag                                                                               \\
                & = \hat J_{\mathrm{s}}(0)-\hat J_{\mathrm{s}}(Ft)+Ft\hat H_0,                                                            \\
    \hat u_2(t) & = \int_0^{Ft}d\Phi_1 \int_0^{\Phi_1} d\Phi_2\; \hat H(\Phi_1) \hat H(\Phi_2) \notag                                     \\
                & = \frac{1}{2}(\hat J_{\mathrm{s}}(0)^2+\hat J_{\mathrm{s}}(Ft)^2)-\hat{J}_{\mathrm{s}}(Ft)\hat{J}_{\mathrm{s}}(0)\notag \\
                & \quad +Ft(\hat H_0 \hat J_{\mathrm{s}}(0)-\hat J_{\mathrm{s}}(Ft)\hat H_0) \notag                                       \\
                & \quad +[(\hat H(Ft)-\hat H(0)),\hat H_0]+\frac{1}{2}(Ft\hat H_0)^2\notag                                                \\
                & \quad  -iFt[H_1,H_{-1}],
\end{align}

\end{document}